\documentclass[twocolumn,showpacs,preprintnumbers,amsmath,amssymb,prl]{revtex4-1}

\usepackage{graphicx}
\usepackage{dcolumn}
\usepackage{bm}
\usepackage{amsmath}
\usepackage{amssymb}
\usepackage{bbold}
\usepackage{siunitx}
\usepackage{color}
\usepackage{braket}
\usepackage{lipsum}
\usepackage{hyperref}
\usepackage{natbib}

\begin{document}

\newcommand{\note}[1]{\textbf{#1}}

\title{Beyond the standard quantum limit of parametric amplification}
\author{\firstname{Michael} \surname{Renger$^{1,2}$}}
\email{michael.renger@wmi.badw.de}
\author{\firstname{S.}~\surname{Pogorzalek$^{1,2}$}}
\author{\firstname{Q.}~\surname{Chen$^{1,2}$}}
\author{\firstname{Y.}~\surname{Nojiri$^{1,2}$}}
\author{\firstname{K.}~\surname{Inomata$^{4,5}$}}
\author{\firstname{Y.}~\surname{Nakamura$^{4,6}$}}
\author{\firstname{M.}~\surname{Partanen$^{1}$}}
\author{\firstname{A.}~\surname{Marx$^{1}$}}
\author{\firstname{R.}~\surname{Gross$^{1,2,3}$}}
\author{\firstname{F.}~\surname{Deppe$^{1,2,3}$}}
\author{\firstname{K. G.}~\surname{Fedorov$^{1,2}$}}
\email{kirill.fedorov@wmi.badw.de}

\affiliation
{
$^{1}$ Walther-Mei{\ss}ner-Institut, Bayerische Akademie der Wissenschaften, D-85748 Garching, Germany \\
$^{2}$ Physik-Department, Technische Universit\"{a}t M\"{u}nchen, D-85748 Garching, Germany \\
$^{3}$ Munich Center for Quantum Science and Technology (MCQST), Schellingstr. 4, 80799 Munich, Germany \\
$^{4}$ RIKEN Center for Emergent Matter Science (CEMS), Wako, Saitama 351-0198, Japan \\
$^{5}$ National Institute of Advanced Industrial Science and Technology, 1-1-1 Umezono, Tsukuba, Ibaraki, 305-8563, Japan \\
$^{6}$ Research Center for Advanced Science and Technology (RCAST), The University of Tokyo, Meguro-ku, Tokyo 153-8904, Japan
}

\date{\today}

\begin{abstract}
The low-noise amplification of weak microwave signals is crucial for countless protocols in quantum information processing. Quantum mechanics sets an ultimate lower limit of half a photon to the added input noise for phase-preserving amplification of narrowband signals, also known as the standard quantum limit (SQL). This limit, which is equivalent to a maximum quantum efficiency of $0.5$, can be overcome by employing nondegenerate parametric amplification of broadband signals. We show that, in principle, a maximum quantum efficiency of 1 can be reached. Experimentally, we find a quantum efficiency of $0.69 \pm 0.02$, well beyond the SQL, by employing a flux-driven Josephson parametric amplifier and broadband thermal signals. We expect that our results allow for fundamental improvements in the detection of ultraweak microwave signals.
\end{abstract}



\maketitle
\begin{figure*}
	\begin{center}
		\includegraphics[width=\linewidth,angle=0,clip]{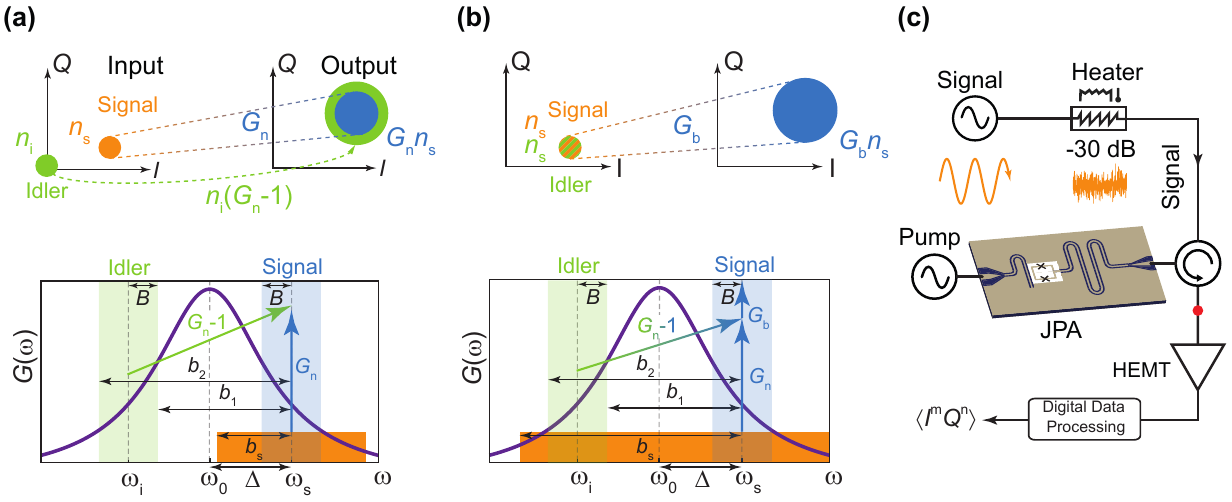}
	\end{center}
	\caption{(a) Top: Phase space transformation for the amplification of narrowband input signals. Colored circles depict the respective variances of the input signal (orange), output signal (blue), and idler mode (green). The narrowband coherent state is amplified with the gain $G_\mathrm{n}$ and the idler mode adds $n_\mathrm{i}(G_\mathrm{n}-1)$ noise photons to the output signal. Bottom: Spectrum of parametric amplification of narrowband signals with a bandwidth $b_\mathrm{s}$. The purple solid line shows a Lorentzian
spectral gain function resulting from the parametric down-conversion process.
For each signal mode, an idler mode at the frequency $2\omega_0 - \omega_\mathrm{s}$ is created. Blue-shaded and green-shaded regions represent measurement bands with full bandwidth $2B$ around the signal and idler modes, respectively. For input signals with $b_\mathrm{s} \leq b_1$, the idler adds at least vacuum fluctuations to the output. (b) Top: Phase space transformation for amplification of broadband input signals. If the input at the idler frequency contributes to the signal, the idler does no longer act as a noise port and the signal is amplified with the total gain $G_\mathrm{b}=2G_\mathrm{n}-1$. Bottom: Spectrum of parametric amplification process for broadband signals. If $b_\mathrm{s} \geq b_2$, each mode in the signal bandwidth corresponds to a correlated input mode on the idler side, resulting in amplification with the total gain $G_\mathrm{b}$ and absence of the SQL. (c) Illustration of the experimental setup. The amplification chain consists of a nondegenerate Josephson parametric amplifier (JPA) and a cryogenic HEMT amplifier. The quadrature moments $\langle I^m Q^n \rangle$ are calculated from the digitized and filtered output signal. The red dot labels the signal reconstruction point.}
	\label{Fig:Fig1}
\end{figure*}

In quantum technology, resolving low power quantum signals in a noisy environment is essential for an efficient readout of quantum states \cite{Schoelkopf_2010}. Linear phase-preserving amplifiers are a central tool to accomplish this task by simultaneously increasing the amplitude in both signal quadratures without altering the signal phase \cite{Caves_2012}. Due to the low energy of microwave photons, this amplification process is of particular relevance for the tomography of quantum microwave states \cite{Lehnert_2011}. State tomography is crucial for experimental protocols with quantum propagating microwaves such as entanglement generation \cite{Menzel_2012,Flurin2012,Fedorov_2016,Fedorov_2018,Schneider_2020}, secure quantum remote state preparation \cite{Pogorzalek_2019}, quantum teleportation \cite{DiCandia_2015}, quantum illumination \cite{LasHeras_2017}, or quantum state transfer \cite{Bienfait_2019}. Furthermore, a conventional dispersive readout of superconducting qubits \cite{Schoelkopf_2004, Goetz_2017, Xie_2018, Goetz_2018, Eddins_2019} relies on the amplification of microwave signals which carry information about the qubit and consist only of a few photons. This approach has proven to be extremely successful and has led to the realization of important milestones in superconducting quantum information processing \cite{Kandala_2017,Arute_2019}. However, fundamental laws of quantum physics imply that any phase-preserving amplifier needs to add at least half a noise photon in the high-gain limit \cite{Caves_1982}. This bound is known as the standard quantum limit (SQL) and results from the bosonic commutation relations of input and output fields constituting the original and amplified signals, respectively. As a result, alternative ways to realize noiseless amplification are important for a large variety of quantum applications which rely on the efficient detection of signal amplitudes such as the parity measurements in multi-qubit systems \cite{Gambetta_2016}, quantum amplitude sensing \cite{Joas_2017}, detection of dark matter axions \cite{Braine_2020}, or the detection of the cosmic microwave background \cite{Braggio_2013}, among others.



In this work, we investigate the nondegenerate parametric amplification of broadband microwave signals and derive conditions under which noiseless amplification is physically possible. This broadband nondegenerate regime is realized by employing a flux-driven Josephson parametric amplifier (JPA) \cite{Yamamoto_2008, Yurke_1989} and is complimentary to the conventional phase-preserving nondegenerate (narrowband and quantum-limited) or phase-sensitive degenerate (narrowband and potentially noiseless) regimes \cite{Zhong_2013,Pogorzalek_2017,Lecocq_2020}. In the current context, the terms ``narrowband'' and ``broadband'' relate to the bandwidth of input signals and not to the bandwidth of the JPA itself. We experimentally demonstrate a JPA operated in the broadband nondegenerate regime achieving linear amplification which is not restricted by the SQL.

\begin{figure}[!t]
	\begin{center}
		\includegraphics[width=0.8\linewidth,angle=0,clip]{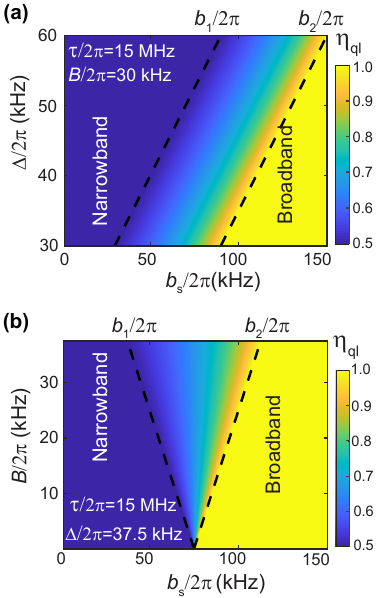}
	\end{center}
	\caption{Limit $\eta_\mathrm{ql}$ of the quantum efficiency $\eta$ as a function of the signal bandwidth $b_\mathrm{s}$ for (a) varying detuning $\Delta$ at a fixed measurement bandwidth $B/2\pi$ = \SI{30}{\kilo \hertz} and (b) for varying measurement bandwidth $B$ at a fixed detuning $\Delta/2\pi = \SI{37.5}{\kilo \hertz}$, according to Eq.\,\,\eqref{Eq:Quantum limit arctan}. For $b_\mathrm{s} \geq b_2$, amplification with $\eta=1$ is possible.}
	\label{Fig:Fig2}
\end{figure}
\begin{figure*}[!t]
	\centering
	\includegraphics[width=\linewidth,angle=0,clip]{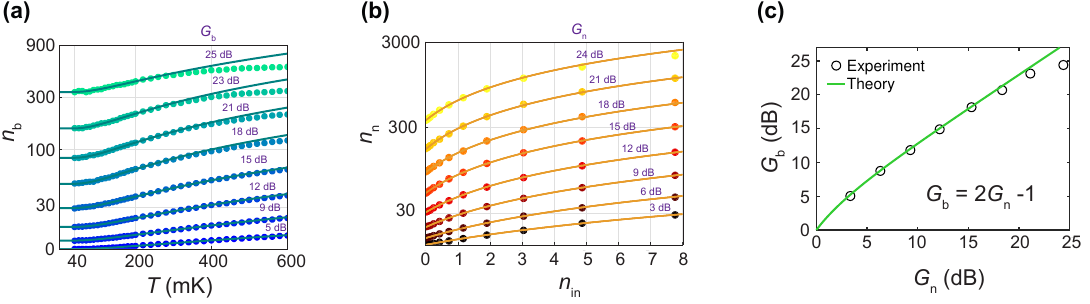}
	\caption{(a) Planck spectroscopy for varying broadband gain $G_\mathrm{b}$. The average photon number $n_\mathrm{b}$ of the amplified signal (dots) is indicated at the reconstruction point. The lines are fits to the corresponding Planck distributions. The offset of each Planck curve is determined by the noise added by the amplification chain. (b) Average photon number $n_\mathrm{n}$ of the amplified signal (dots) at the reconstruction point for the amplification of a coherent signal for different values of the narrowband gain $G_\mathrm{n}$. The average photon number of the input signal is labeled with $n_\mathrm{in}$. From the offsets of the linear fits (lines), we can calculate the number of added noise photons. (c) Experimentally determined broadband gain $G_\mathrm{b}$ vs. narrowband gain $G_\mathrm{n}$ of the JPA.}
	\label{Fig:Fig3}
\end{figure*}
An ideal linear amplifier increases the signal photon number $n_\mathrm{s}$ by the power gain factor $G_\mathrm{n}$ \cite{Caves_2012}. The fluctuations in the output signal consist of the amplified vacuum fluctuations of the input signal and the noise photons $n_\mathrm{f}$, added by the amplifier, where $n_\mathrm{f}$ is referred to the amplifier input. We use the quantum efficiency $\eta$ to characterize the noise performance of our amplifiers \cite{Blais_2017}. The quantum efficiency is defined as the ratio between vacuum fluctuations in the input signal and fluctuations in the output signal. Thus, $\eta$ can be expressed as
\begin{equation}\label{Eq:1_QuantumEfficiency}
\eta = \frac{1}{1+2n_\mathrm{f}}.
\end{equation}
Parametrically driving the JPA results in amplification of the incoming signal and creation of an additional phase-conjugated idler mode. This idler mode consists of, at least, vacuum fluctuations, implying that it carries $n_\mathrm{i}\geq 1/2$ photons. As a result of the narrowband parametric signal amplification, the idler adds \cite{Devoret_2016}
\begin{equation}\label{Eq:2_SQLphotons}
n_\mathrm{f} =n_\mathrm{i} \frac{G_\mathrm{n}-1}{G_\mathrm{n}} \geq \frac{1}{2}\left(1-\frac{1}{G_\mathrm{n}}\right)
\end{equation}
noise photons to the signal, referred to the amplifier input, as schematically depicted in Fig.\,\ref{Fig:Fig1}(a). Equation \eqref{Eq:2_SQLphotons} implies that $\eta$ is bounded by
\begin{equation}\label{Eq:3_SQLefficiency}
\eta \leq \frac{G_\mathrm{n}}{2G_\mathrm{n}-1},
\end{equation}
reaching $1/2$ in the high-gain limit, $G_\mathrm{n} \gg 1$. Figure\,\ref{Fig:Fig1}(b) illustrates the amplification process for broadband signals, where the input signal bandwidth is large enough to cover both the signal and idler modes of the parametric amplifier. As a result, the idler mode does not add any noise but contributes to the amplified output signal. Thus, we expect that a quantum efficiency of $\eta=1$ can be reached in this case.  
Parametric amplification can be realized by driving a nonlinear electromagnetic resonator with a strong coherent field (pump) at $\omega_\mathrm{p}=2\omega_0$, where  $\omega_0$ is the resonance frequency \cite{Pogorzalek_2017, Devoret_2016}. In the resulting three-wave mixing process, a pump photon splits into a signal photon at frequency $\omega_\mathrm{s}=\omega_\mathrm{p}/2+\Delta$ and a corresponding idler photon at $\omega_\mathrm{i}=\omega_\mathrm{p}/2-\Delta$. Here, $\Delta$ denotes the detuning of the signal reconstruction frequency $\omega_\mathrm{s}$ from $\omega_0$. In the nondegenerate regime, we have $\Delta \neq 0$, leading to separated signal and idler modes in frequency space \cite{Yamamoto_2016}. This parametric down-conversion process results in a Lorentzian spectral gain function, which is depicted by the purple solid line in Fig.\,\ref{Fig:Fig1}(a), bottom. The signal is detected within the measurement bandwidth $2B$ around $\omega_\mathrm{s}$. For a narrowband coherent state, the idler necessarily adds broadband noise to the signal, leading to broadening of the output variances (see Fig.\,1(a), top). In contrast, if the signal bandwidth $b_\mathrm{s}$ is large enough to cover the idler input modes, as shown in Fig.\,\ref{Fig:Fig1}(b), the amplified output signal consists of contributions from both the signal and idler modes. In this case, the idler no longer serves as a noise port but rather as an additional signal port. The part of the input signal at the idler frequency is mixed into the measurement bandwidth with the gain $G_\mathrm{n}-1$, yielding a total broadband power gain
\begin{equation}\label{Eq: broadband gain}
G_\mathrm{b} = 2G_\mathrm{n}-1.
\end{equation}
The SQL fundamentally results from the fact that the amplifier input and output signals have to fulfill the bosonic commutation relations. We assume that the JPA output $\hat{c}(\omega)$ results from a linear combination of amplified incoming signals $\hat{a}(\tilde{\omega})$ and phase-conjugated idler signals $\hat{a}^\dagger(\tilde{{\omega}}_\mathrm{i})$ as well as from an additive noise mode $\hat{f}(\omega)$ \cite{Caves_1982}
\begin{equation}\label{Eq:Input}
\hat{c}(\omega)=\int_{\mathcal{I}}d\tilde{\omega}\left[M(\omega, \tilde{\omega})\hat{a}(\tilde{\omega}) +  L(\omega, \tilde{\omega})\hat{a}^\dagger(\tilde{\omega}) \right] + \hat{f}(\omega),
\end{equation}
where the integral is taken over all input modes $\mathcal{I}$. Frequencies $\tilde{\omega}$ denote input modes, whereas output modes are described by $\omega$. We assume that the input signal is centered around the frequency $\omega_\mathrm{s}$ and has a single-side bandwidth $b_\mathrm{s}$ (total bandwidth $2b_\mathrm{s}$), resulting in $\mathcal{I}\,=\,[\omega_\mathrm{s}-b_\mathrm{s}, \omega_\mathrm{s}+b_\mathrm{s}]$. The signal amplitude gain $M(\omega, \tilde{\omega})$ and the idler amplitude gain $L(\omega, \tilde{\omega})$ satisfy $M(\omega, \tilde{\omega}) = M(\tilde{\omega}) \delta(\omega- \tilde{\omega})$ and $L(\omega, \tilde{\omega}) = L(\tilde{{\omega}}_\mathrm{i}) \delta(\omega-\tilde{{\omega}}_\mathrm{i})$, respectively, where $\tilde{\omega}_\mathrm{i}=2\omega_0-\tilde{\omega}$. Furthermore, $|M(\omega)|^2 = G_\mathrm{n}(\omega)$ and $|L(\omega_\mathrm{i})|^2 = G_\mathrm{n}(\omega)-1$, where $G_\mathrm{n}(\omega)$ is the power gain at frequency $\omega$. The strength of the additive noise is described with the noise power spectral density $S_\mathrm{f}(\omega)$ of the bosonic mode $\hat{f}(\omega)$, corresponding to the number of noise photons per bandwidth. We evaluate the integral in Eq.\,\eqref{Eq:Input} and calculate the commutator $[\hat{c}(\omega), \hat{c}^\dagger(\omega^\prime)]$. Next, we calculate the spectral autocorrelation function of the fluctuations $\langle |\Delta \hat{f}(\omega, \omega^\prime)|^2\rangle = 2\pi G_\mathrm{n}(\omega) S_\mathrm{f}(\omega)\delta(\omega - \omega^\prime)$. This allows us to use the bosonic continuum commutation relations and the Heisenberg uncertainty principle $\langle |\Delta \hat{f}(\omega, \omega^\prime)|^2 \rangle \geq 1/2|\langle [\hat{f}(\omega), \hat{f}^\dagger(\omega^\prime)] \rangle|$ to show that $S_\mathrm{f}(\omega)$ satisfies
\begin{align}\label{generalizedCaves} \notag
& \int_{\omega_\mathrm{s}-b_\mathrm{s}}^{\omega_\mathrm{s}+b_\mathrm{s}}S_\mathrm{f}(\omega)d\omega = 2b_\mathrm{s} n_\mathrm{f} \geq 2b_\mathrm{s}n_\mathrm{ql} = \\& = \frac{1}{2}  \int_{\omega_\mathrm{s}-b_\mathrm{s}}^{\omega_\mathrm{s}+b_\mathrm{s}}\left[1- \mathbb{1}_{\mathrm{i}}(\omega) + \frac{1}{G_\mathrm{n}(\omega)}(  \mathbb{1}_\mathrm{i}(\omega)-1) \right] d\omega, \end{align}
where $n_\mathrm{ql}$ is the quantum limit for the number of additive noise photons \cite{Supplement}. We define
\begin{equation}
\mathbb{1}_{\mathrm{i}}(\omega) = \Theta(\omega - 2\omega_0+\omega_\mathrm{s}+b_\mathrm{s})-\Theta(\omega- 2\omega_0+\omega_\mathrm{s}-b_\mathrm{s}),
\end{equation}
where $\Theta$ is the Heaviside step function. In Eq.\,\eqref{generalizedCaves}, the range of integration over $\omega$ is limited by $b_\mathrm{s}$, if $b_\mathrm{s} < B$, otherwise it is set by $B$. We observe that there are two threshold values of the bandwidth: $b_1 = 2\Delta -B$ and $b_2 = 2\Delta + B$. For $b_\mathrm{s} \leq b_1$, Eq.\,\eqref{generalizedCaves} reproduces the SQL, whereas for $b_1\,\leq b_\mathrm{s}\,\leq\,b_2$, the input signal starts to overlap with the idler modes and the lower bound for the additive noise decreases. In the broadband case $b_\mathrm{s} \geq b_2$, the signal covers all idler modes and Eq.\,\eqref{generalizedCaves} reduces to $n_\mathrm{f} \geq 0$, implying that there is no fundamental lower limit for the added noise.
Our experimental setup is schematically shown in Fig.\,\ref{Fig:Fig1}(c) and consists of a flux-driven JPA serially connected to a cryogenic high-electron-mobility transistor (HEMT) amplifier with a gain of $G_\mathrm{H}=\SI{41}{\decibel}$. The JPA is operated in the nondegenerate regime, which is realized by detuning the signal frequency $\omega_\mathrm{s}$ by $\Delta/2\pi =\SI{300}{\kilo \hertz}$ from half the pump frequency $\omega_\mathrm{p}/2=\omega_0$. A circulator at the JPA input separates the resonator input and output fields. The moments of the output signal are reconstructed with a bandwidth $B/2\pi=\SI{200}{\kilo \hertz}$ using the reference-state reconstruction method at the reconstruction point indicated by the red circle in Fig.\,\ref{Fig:Fig1}(c) \cite{Fedorov_2018, Eichler_2011}. The experiment is performed with two distinct JPAs, labelled JPA\,1 and JPA\,2, which are operated at different flux spots to check for reproducibility of our results. For JPA\,1 (JPA\,2), we reconstruct the signal at $\omega_\mathrm{s}/2\pi=\SI{5.500}{\giga \hertz}$ ($\SI{5.435}{\giga \hertz}$). A coherent tone can be applied via a microwave input line and a heatable $\SI{30}{\decibel}$ attenuator allows us to generate thermal states as broadband input signals.
We solve Eq.\,\eqref{generalizedCaves} for the Lorentzian JPA gain function $G(\omega)=1+ G_0 b_\mathrm{J}^2/(b_\mathrm{J}^2+(\omega-\omega_0)^2)$, where $G_0$ denotes the maximal JPA gain and $b_\mathrm{J}$ is the half width at half maximum JPA bandwidth \cite{Yamamoto_2016}. Then, assuming $G_0 \gg 1$, the quantum limit $n_\mathrm{ql}$ for the number of additive noise photons is given by
\begin{equation}\label{Eq:Quantum limit arctan}
n_\mathrm{ql} = \frac{1}{4\beta} \begin{cases}
\frac{\beta}{\beta_\mathrm{s}}\arctan\left(\frac{2\beta_\mathrm{s}}{1+\delta^2-\beta_\mathrm{s}^2}\right) & b_\mathrm{s}\leq B, \\
\arctan\left(\frac{2\beta}{1+\delta^2-\beta^2}\right) &  B  \leq b_\mathrm{s} \leq b_1, \\
\arctan\left(\frac{2(\delta+\beta-\beta_\mathrm{s})}{1 + \beta \beta_\mathrm{s} + \delta (\beta_\mathrm{s}-\beta)-\delta^2}\right) & b_1 \leq b_\mathrm{s} \leq b_2, \\
0 & b_\mathrm{s}\geq b_2,
\end{cases}
\end{equation}
with $\beta = \mathrm{B}/\tau$, $\beta_\mathrm{s} = b_\mathrm{s}/\tau$, and $\delta = \Delta/\tau$, where $\tau \equiv b_\mathrm{J} \sqrt{G_0}$ denotes the gain-bandwidth product \cite{Supplement, Zhong_2013}. The corresponding limit $\eta_\mathrm{ql}$ of the quantum efficiency $\eta$ can be calculated with Eq.\,\eqref{Eq:1_QuantumEfficiency}. The solution Eq.\,\eqref{Eq:Quantum limit arctan} allows us to distinguish quantitatively between broadband and narrowband regimes and is plotted for $\tau/2\pi \simeq \SI{15}{\mega \hertz}$ in Fig.\,\ref{Fig:Fig2}(a) for varying $\Delta$ with $B/2\pi=\SI{30}{\kilo \hertz}$ and in Fig.\,\ref{Fig:Fig2}(b) for varying $B$ with $\Delta/2\pi=\SI{37.5}{\kilo \hertz}$.
According to Eq.\,\eqref{Eq:Quantum limit arctan}, we obtain $\eta_\mathrm{ql} \simeq 1/2$ for coherent input signals, approximately reproducing Eq.\,\eqref{Eq:3_SQLefficiency}, whereas we expect $\eta_\mathrm{ql}=1$ for broadband signals. 
We experimentally extract the quantum efficiency by measuring the total noise photon number of the amplification chain. To achieve this goal, we vary the temperature of the heatable $\SI{30}{\decibel}$ attenuator from $\SI{40}{\milli \kelvin}$ to $\SI{600}{\milli \kelvin}$ and perform Planck spectroscopy of the amplification chain \cite{Gross_2010}. As a result, we detect the photon number $n_\mathrm{b}$ at the reconstruction point for varying broadband gain $G_\mathrm{b}$ and show the result of this measurement in Fig.\,\ref{Fig:Fig3}(a) for JPA\,2. For each value of $G_\mathrm{b}$, the experimentally determined outcomes for $n_\mathrm{b}$ (dots) are fitted with corresponding Planck distributions (cyan solid lines) and the respective noise photon number is extracted from the offset.

The quantum efficiency for narrowband signals is determined in a similar experiment by amplifying a coherent input tone with varying input photon number $n_\mathrm{in}$ for different JPA gains $G_\mathrm{n}$. For each value of $G_\mathrm{n}$, the amplifier response $n_\mathrm{n}$ at the reconstruction point is linearly fitted, which allows us to extract the noise photons from the respective offset. This procedure also proves that the JPA acts as a linear amplifier here. Figure\,\ref{Fig:Fig3}(b) shows a logarithmic plot of the experimental results (dots) for JPA\,2 as well as the respective linear fits (orange lines).
The dependence of the broadband gain $G_\mathrm{b}$ on $G_\mathrm{n}$ is depicted in Fig.\,\ref{Fig:Fig3}(c). The results are in agreement with Eq.\,\eqref{Eq: broadband gain}.

\begin{figure}
	\begin{center}
		\includegraphics[width=0.8\columnwidth,angle=0,clip]{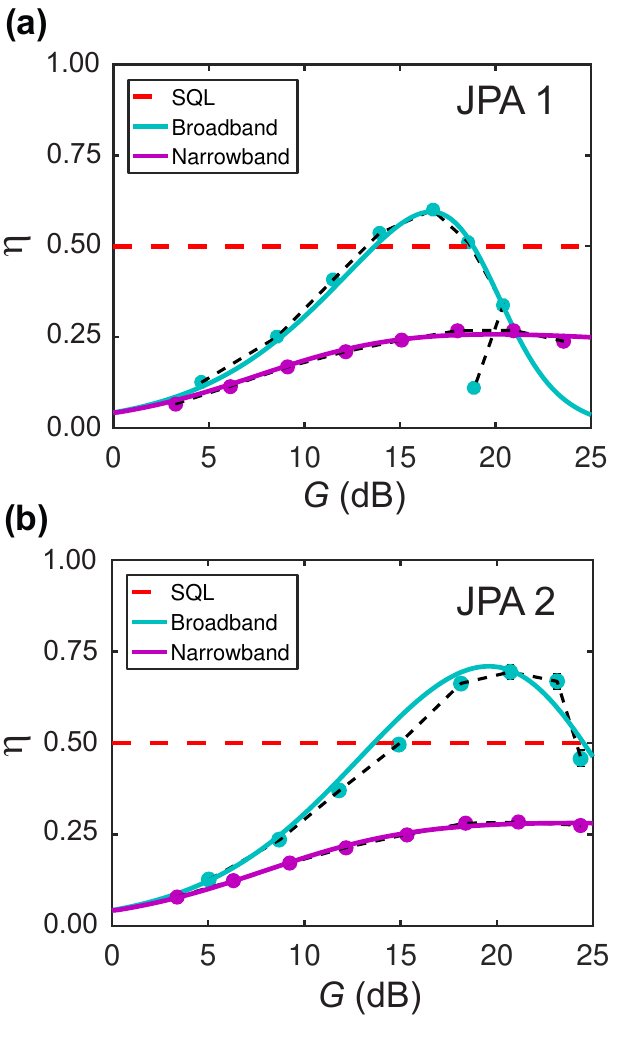}
	\end{center}
	\caption{Experimental quantum efficiency of the flux-driven JPAs for broadband (cyan dots) and narrowband (purple dots) input signals. Panel (a) shows the data for JPA 1 and panel (b) the data for JPA 2. The red dashed line represents the SQL. The gain dependence of the quantum efficiencies for the narrowband and broadband cases is fitted using Eq.\,\eqref{Eq:PowerLaw}.}
	\label{Fig:Fig4}
\end{figure}

In Fig.\,\ref{Fig:Fig4}(a), we plot the measured quantum efficiencies for JPA\,1 for the amplification of thermal states (cyan dots) and coherent states (purple dots), respectively. The red dashed line depicts the SQL determined by Eq.\,\eqref{Eq:3_SQLefficiency}. Figure\,\ref{Fig:Fig4}(b) shows the result for the same experiment with JPA\,2 instead of JPA\,1. For both JPAs, we find a gain region where we clearly exceed the SQL for the amplification of broadband states. 

Importantly, Fig.\,\ref{Fig:Fig4}(b) shows that we can achieve a maximal quantum efficiency $\eta=0.69 \pm 0.02$ with our setup, which substantially exceeds the SQL.
The deviation from the theoretically achievable quantum efficiency of unity can be explained by noise in the pump signal, as discussed below. Furthermore, we observe that the experimentally determined dependence of quantum efficiency on the gain in Fig. \ref{Fig:Fig4} reaches a maximum and decreases for high gains for the narrowband and broadband case. For low JPA gains, the noise photons $n_\mathrm{H}=11.3$, which are added by the HEMT, limit the quantum efficiency. This contribution becomes irrelevant in the high-gain limit, since its influence decreases with $1/G$, where $G=G_\mathrm{n}$ ($G=G_\mathrm{b}$) for narrowband (broadband) amplification. Since the parametric gain depends on the pump power, fluctuations in the pump photon number imply additional noise in the signal mode \cite{Kylemark_2006}. We describe the noisy pump signal in the frame rotating at a pump frequency $\omega_\mathrm{p}$ by
\begin{equation}
\hat{A}(t) = (\alpha_0 + \hat{f}_\mathrm{p}(t))e^{-i\omega_\mathrm{p}t},
\end{equation}
where $\alpha_0$ denotes the amplitude of the coherent pump signal and the operator $\hat{f}_\mathrm{p}(t)$ represents uncorrelated white noise. We assume that $\hat{f}_\mathrm{p}(t)$ obeys thermal statistics and calculate the variance  $\sigma_\mathrm{p}^2$ of the corresponding power fluctuations by applying the Wiener-Khinchine theorem \cite{Supplement, Olsson_1989}. We find
\begin{equation}
\sigma_\mathrm{p}^2 = n_\mathrm{p}(2n_\mathrm{th}+1)+2n_\mathrm{th}^2+n_\mathrm{th},
\end{equation}
where $n_\mathrm{p}$ is the average pump photon number and $n_\mathrm{th}=\langle \hat{f}_\mathrm{p}^\dagger(t)\hat{f}_\mathrm{p}(t) \rangle$ is the number of thermal noise photons. The experimentally determined dependence of the parametric gain on the pump power can be fitted by an exponential function \cite{Supplement}. The gain-dependent JPA noise $n_\mathrm{J}(G)$ can then be approximated by
\begin{equation}\label{Eq:PowerLaw}
n_\mathrm{J}(G) = n_\mathrm{J}^\prime (G-1)^\epsilon,
\end{equation}
where the exponent $\epsilon$ depends on JPA parameters and $n_\mathrm{J}^\prime$ is a constant prefactor \cite{Supplement}.  We use Eq.\,\eqref{Eq:PowerLaw} to fit the measured quantum efficiencies in Figs. \ref{Fig:Fig4}(a) and (b) and treat $n_\mathrm{J}^\prime$ and $\epsilon$ as fitting parameters \cite{Supplement}. In Fig.\,\ref{Fig:Fig4}(a), the last data point for broadband amplification is not considered for the fit, as JPA\,1 starts entering a nonlinear compression regime. The fit is depicted by the solid lines in Figs. \ref{Fig:Fig4}(a) and (b) for both JPAs and successfully reproduces the maximum as well as the behavior for low gain values.

Noiseless parametric amplification of broadband signals can be exploited for various applications. For instance, it can be used for high-efficiency parity detection of entangled superconducting qubits via readout of dispersively coupled resonators. Assuming that these resonators are probed at the respective signal and idler frequencies of a readout JPA, one can amplify the combined resonator response with quantum efficiency beyond the SQL. Thus, one can implement a parity readout with, potentially, unit efficiency by combining the two-mode probe via nondegenerate parametric amplification. Another possible application could be a direct broadband dispersive qubit readout with weak thermal states generated by a heatable attenuator \cite{Goetz_2017} or naturally occurring thermal states due to residual heating. In this case, propagating weak thermal states can be used to probe the readout resonator, which is then efficiently amplified by the flux-driven JPA with quantum efficiency beyond the SQL. 

In conclusion, we have investigated a nondegenerate linear parametric amplification of broadband signals and have derived a quantitative criterion for an input signal bandwidth under which a quantum efficiency of $\eta=1$ can be achieved. We have used a superconducting flux-driven JPA to experimentally determine the quantum efficiencies for amplification of broadband thermal states and demonstrated $\eta=0.69\pm 0.02$ which significantly exceeds the SQL $\eta_\mathrm{ql}=0.5$. Thus, we have verified that for the parametric amplification of broadband input states, an idler mode may also carry signal information and does not add extra noise to the output. Furthermore, we have shown that the gain dependence of $\eta$ can be explained by the photon number fluctuations in the pump tone. The fact that the quantum efficiency for broadband signal amplification does not possess a corresponding quantum limit can be exploited in experiments where extremely low noise amplification is a key prerequisite. We envision important applications of our results in the field of quantum information processing, in particular for qubit readout, parity detection, and quantum amplitude sensing.

\noindent
\textbf{\large Acknowledgments}

\noindent
We acknowledge support by the German Research Foundation through the Munich Center for Quantum Science and Technology (MCQST), Elite Network of Bavaria through the program ExQM, EU Flagship project QMiCS (Grant No. 820505).
\medskip\noindent

\medskip\noindent
\textbf{\large Author contributions}

\noindent
K.G.F. and F.D. planned the experiment. M.R., S.P., and K.G.F. performed the measurements and analyzed the data. M.R. and K.G.F. developed the theory. Q.C., Y.N., and M.P. contributed to development of the measurement software and experimental set-up. K.I. and Y.N. provided the JPA samples. F.D., A.M., and R.G. supervised the experimental part of this work. M.R. and K.G.F. wrote the manuscript. All authors contributed to discussions and proofreading of the manuscript.

\textbf{\large Competing financial interests}

\noindent
The authors declare no competing financial interests.

\medskip\noindent
\textbf{\large Data availability}

\noindent
The data that support the findings of this study are available from the corresponding author upon reasonable request.
\bibliography{SQL_Paper}
\medskip\noindent

\clearpage

\end{document}